# $CeMnNi_4$: A soft ferromagnet with a high degree of transport spin polarization

Surjeet Singh, Goutam Sheet, Pratap Raychaudhuri, Sudesh Kumar Dhar
*Department of Condensed Matter Physics and Materials Science*
*Tata Institute of Fundamental Research*
*Homi Bhabha Rd., Colaba, Mumbai 400005, India.*

**Abstract:** In this letter we introduce a new soft ferromagnetic compound, i.e. $CeMnNi_4$, which exhibits a large moment (~4.95$\mu_B$/Mn) and high degree of spin polarization. The system has a ferromagnetic transition temperature of 148K. Isothermal magnetization measurements at 5K reveal that the material is a soft ferromagnet with a magnetization saturating at about 500Oe and a coercive field of < 5 Oe. We determine the transport spin polarization of this material from Point Contact Andreev Reflection measurements to be 66% thereby making this material potentially important for spintronic applications.

With the advent of "spintronics", realizing the ability to inject and manipulate spin polarized electrons in metals and semiconductors has gained particular relevance[1]. The basic philosophy behind this emerging area of research is to be able to inject spin polarized electrons and manipulate the spin as well as the charge to realize novel device schemes and to enhance functionality of existing devices. In this context the search for ferromagnets with a high degree of spin polarization, which can be used as source of spin polarized electrons, has gained particular importance. The forerunner in this class of materials[2] is $CrO_2$ which exhibits a near complete spin polarization (96%) at the Fermi level. However, difficulties in synthesizing high quality $CrO_2$ films through conventional deposition techniques have so far hindered the use of this material for spin injection in devices. Though several other compounds such as $CoS_2$[3], the Heusler alloy NiMnSb[4] and $Co_2MnSi$[5] and the doped rare-earth manganite $La_{0.7}Sr_{0.3}MnO_3$[6] have been predicted to have a large degree of spin polarization at the Fermi level from electronic structure calculations, experimental observations of spin polarization exceeding 60% have been very few. Several detrimental factors, such as depolarization of the electrons at a disordered interface[7], spin orbit coupling and spin fluctuation effects are believed to be responsible for the reduced spin polarization observed in these compounds. It is also important to note that though the degree of spin polarization is one of the most important criterion for a spin source, several other factors are also crucial to realize efficient spin injection: These include, the Fermi velocity mismatch between the source and the material in which the electrons are injected, the ease to fabricate the material in thin film form, the robustness of the spin polarization against detrimental effects of disorder and the magnetic properties of the material. Since these requirements are expected to be different for different applications, identifying ferromagnets with a large degree of spin polarization is crucial for the progress of spintronics.

In this letter we introduce a new ferromagnetic material, namely, $CeMnNi_4$, with a $T_c$~140K which exhibits a large degree of transport spin polarization[8]. This material is derived from the Pauli paramagnet $CeNi_5$[9], by substituting one Ni-site with Mn in the parent compound. Magnetic measurements on $CeMnNi_4$ show that this material is a soft ferromagnet with a technical saturation at about 4kOe and a coercive field of less than 5 Oe. Point Contact Andreev Reflection (PCAR) measurements on this compound reveal that the material has a large degree of transport spin polarization.

The alloy $CeMnNi_4$ was prepared by arc-melting the constituents in several steps. In the first step, stoichiometric amounts of Ce and Ni were melted to form Ce+4Ni alloy. Subsequently, small amounts of Mn (~20mg) is added to the alloy obtained from the previous melt and the alloy is re-melted till until the required amount of Mn is reacted to obtain $CeMnNi_4$. In each step the weight of the alloy is monitored and any weight loss due to the volatile Mn is compensated by taking extra Mn in the subsequent melting. The alloy thus obtained is then re-melted several times to ensure proper homogenization. The as cast button was then sealed in an evacuated quartz tube and heated to $700^0$C at $1.5^0$C/min and kept at $700^0$C for 2 days. The temperature was then gently raised to $800^0$C and maintained for 8 days followed by quenching in liquid nitrogen. Thee sample was characterized by X-ray diffraction using Cu-$K_\alpha$ radiation. Magnetic measurements were carried out on a SQUID magnetometer. The transport spin polarization was measured using a home built PCAR setup operating down to 2.8K in a continuous flow cryostat. Prior to the measurements the sample surface was polished with fine emory paper to a mirror finish and immediately loaded for measurements to avoid surface degradation. Mechanically cut fine tips of Nb were engaged on the sample surface at low temperature using a differential screw arrangement to establish a point contact. The differential conductance (G(V)) of the contact was recorded as a function of voltage (V) using a 4-probe modulation technique operating at 372Hz. The resistance of all the point contacts analyzed in this paper was between 10-20Ω.

Figure 1 shows the powder x-ray diffraction (XRD) pattern of $CeMnNi_4$. All the peaks could be indexed on the basis of cubic $AuBe_5$ (C15b)-type structure with a lattice parameter *a*=6.987Å. Our result is in conformity with the earlier observation by Kalychak et al.[10] who had found a change of structure in $CeMn_xNi_{5-x}$ from hexagonal $CaCu_5$ ($x \leq 0.9$) to cubic $AuBe_5$-type at higher Mn concentrations; i.e $0.9 \leq x \leq 2.1$. In analogy with $UMNi_4$ and $ZrMNi_4$[11] (M = In, Zn and Sn), a possible superlattice cubic structure of $MgSnCu_4$-type for $CeMnNi_4$ is inferred from the presence of superstructure lines (200) and (420) in the x-ray diffraction spectrum. The superstructure arises when the Mn atoms selectively substitute for the Ni atoms at the 4(c) site of the cubic $AuBe_5$-type unit cell.

Magnetisation versus temperature (M-T) for the $CeMnNi_4$ sample recorded in a field of 100Oe (Figure 2(a)) reveals a sharp ferromagnetic transition with $T_c$~140K. The signature of the ferromagnetic transition is also seen as a pronounced knee in the resistivity vs. temperature (ρ-T) at the same temperature (*inset* Fig. 2(a))[12]. A saturation magnetization of $4.94\mu_B$/f.u is obtained from isothermal magnetization versus field (M-H) scan recorded at 5K (Fig. 2(b)). While in-principle, Ce, Mn and Ni can all possess magnetic moments, the parent compound $CeNi_5$ is a Pauli paramagnet where Ce and Ni do not carry any moment[13]. We therefore believe that the ferromagnetic transition at 140K arises due to the ordering of the Mn moments, which in-turn implies that manganese in this compound is in a high spin $Mn^{2+}$ valence state. The 4 quadrant M-H scan also reveals that the system has very low hysteresis (*inset* Fig.2(b)) with a coercive field $H_c$<5Oe and a very low saturation field ~500Oe. The observation of such a small hysteresis and saturating field for a strong ferromagnet is intriguing from the point of view of application since the magnetization can be reversed by the application of small magnetic field. The compound shows a metallic behavior in the entire temperature range (*inset* Fig. 2(a)) with a residual resistivity, $\rho_0$~240μΩ-cm.

In order to explore the transport spin polarization at the Fermi level we carried out PCAR measurements using a superconducting Nb tip. Figure 3(a) shows the representative PCAR G(V)-V spectra recorded at 2.8K. The spectra are fitted with the Blonder-Tinkham-Klapzijk (BTK) theory[14], modified to take into account the effect of spin polarization in the ferromagnet[15]. All the spectra were fitted using three fitting parameters: The superconducting energy gap Δ, the transport spin polarization $P_t$ and the barrier parameter Z, which characterizes the potential barrier at the interface. The extracted value of $P_t$ shows a decrease with increasing value of Z (Fig. 3(b)), a feature which has been observed for a wide variety of ferromagnets. This decrease is believed to arise from the spin depolarization at the magnetically disordered scattering barrier formed at the interface[16,17]. The intrinsic value of $P_t$ is therefore extracted by linearly extrapolating the $P_t$ vs. Z curve to Z=0. The intrinsic value of the transport spin polarization extracted in this way is $P_t$ ~ 66%.

It is important to note that unlike spin polarized photoemission which measures the spin polarization in the density of states (DOS), i.e. $P = \frac{N_\uparrow - N_\downarrow}{N_\uparrow + N_\downarrow}$, PCAR measures the spin polarization in the transport current. This quantity varies depending on whether the contact is in the ballistic regime where the contact diameter ($d$) is smaller than both the elastic($l_e$) and the inelastic($l_i$) mean free paths, or the diffusive regime where $l_i<d<l_e$. The transport spin polarization is given by $P_t=\frac{\langle N_\uparrow v_{F\uparrow}\rangle_{FS}-\langle N_\downarrow v_{F\downarrow}\rangle_{FS}}{\langle N_\uparrow v_{F\uparrow}\rangle_{FS}+\langle N_\downarrow v_{F\downarrow}\rangle_{FS}}$ for a ballistic contact and $P_t=\frac{\langle N_\uparrow v_{F\uparrow}^2\rangle_{FS}-\langle N_\downarrow v_{F\downarrow}^2\rangle_{FS}}{\langle N_\uparrow v_{F\uparrow}^2\rangle_{FS}+\langle N_\downarrow v_{F\downarrow}^2\rangle_{FS}}$ for a diffusive contact, where $N\uparrow$ and $N\downarrow$ are the DOS of the up and down spin channels at Fermi level and $v_{F\uparrow}$ and $v_{F\downarrow}$ are the respective Fermi velocities[18]. The latter characterizes the spin polarization of a bulk current in the ferromagnet as well as the current through a low transparency barrier. Though, at present, we do not have an estimate of $l_i$ and $l_e$ in this material, from the relatively large residual resistivity (240μΩ-cm) it is likely that the contact is in the diffusive limit. However, in the absence of a knowledge on the up and down Fermi velocities, we cannot say whether the large spin polarization arises predominantly due to mismatch in Fermi velocities of the up and down spin electrons or the difference in the up and down DOS.

In summary, we have synthesized a new soft ferromagnet with very low hysteresis exhibiting a large moment and a high value of transport spin polarization by doping the Pauli paramagnet $CeNi_5$ with manganese. This material could be potentially important in devices where switching of the spin at low magnetic field is desired. We believe that it would be interesting to carry out detailed band structure calculation to establish the origin of large spin polarization in this compound.

*Acknowledgements:* The authors would like to thank Subhash Pai, Vivas Bagwe and Ruta N. Kulkarni for technical support. GS would like to thank TIFR Endowment Fund for partial financial support.

**Figure captions**

Figure 1. Powder X-ray diffraction spectrum of $CeMnNi_4$. The peaks are indexed on the basis of C15b structure. The lines marked "*" are the superstructure lines arising from the ordering of Ce and Mn.

Figure 2. (a) Magnetization versus temperature of $CeMnNi_4$ measured in a field of 100Oe; the *inset* shows the resistivity as a function of temperature. (b) Magnetization versus field (M-H) measured at 5K; the *inset* shows a blow up of the M-H loop close to zero field.

Figure 3. Representative PCAR, G-V spectra measured at 3.4K using a Nb tip. The filled circles show the experimental data and the solid lines show the best fit to the modified BTK model. The best fit values of $P_t$, Z and $\Delta$ are shown in the figure. The values of $\Delta$ extracted from the best fit curves of all the spectra range between 1.3-1.4meV.

Figure 4. Variation of the extracted value of $P_t$ as a function of barrier height parameter Z. The solid line shows the linear extrapolation giving an intrinsic transport spin polarization $P_t \approx 66.5\%$.

[12] The temperature dependence of the resistivity is however unusual: The resistivity decreases very slowly from 300K to 140K, below which it decreases rapidly down to the lowest temperature with a residual resistance ratio, $\frac{\rho(T=300K)}{\rho(T=4.2K)}\sim 1.75$. At present we do not have a clear explanation for this unusual temperature dependence.

[13] A further evidence that Ce atoms do not carry any moment is a absence of a second transition down to 1.5K. If Ce in $CeMnNi_4$ had a stable moment a second transition arising from the ordering of the Ce moments due to RKKY exchange interaction should be observable at low temperatures.

[17] Alternatively, it has also been suggested that the dependence of $P_t$ on Z is an outcome of wrongly estimating the superconducting energy gap while fitting the PCAR spectra with a ballistic theory. In that case however, one expects a systematic variation of $P_t$ with the best fit values of $\Delta$, a feature that we do not observe in our experiments. For more details, see, G. T. Woods et al., Phys. Rev. B **70**, 054416 (2004).

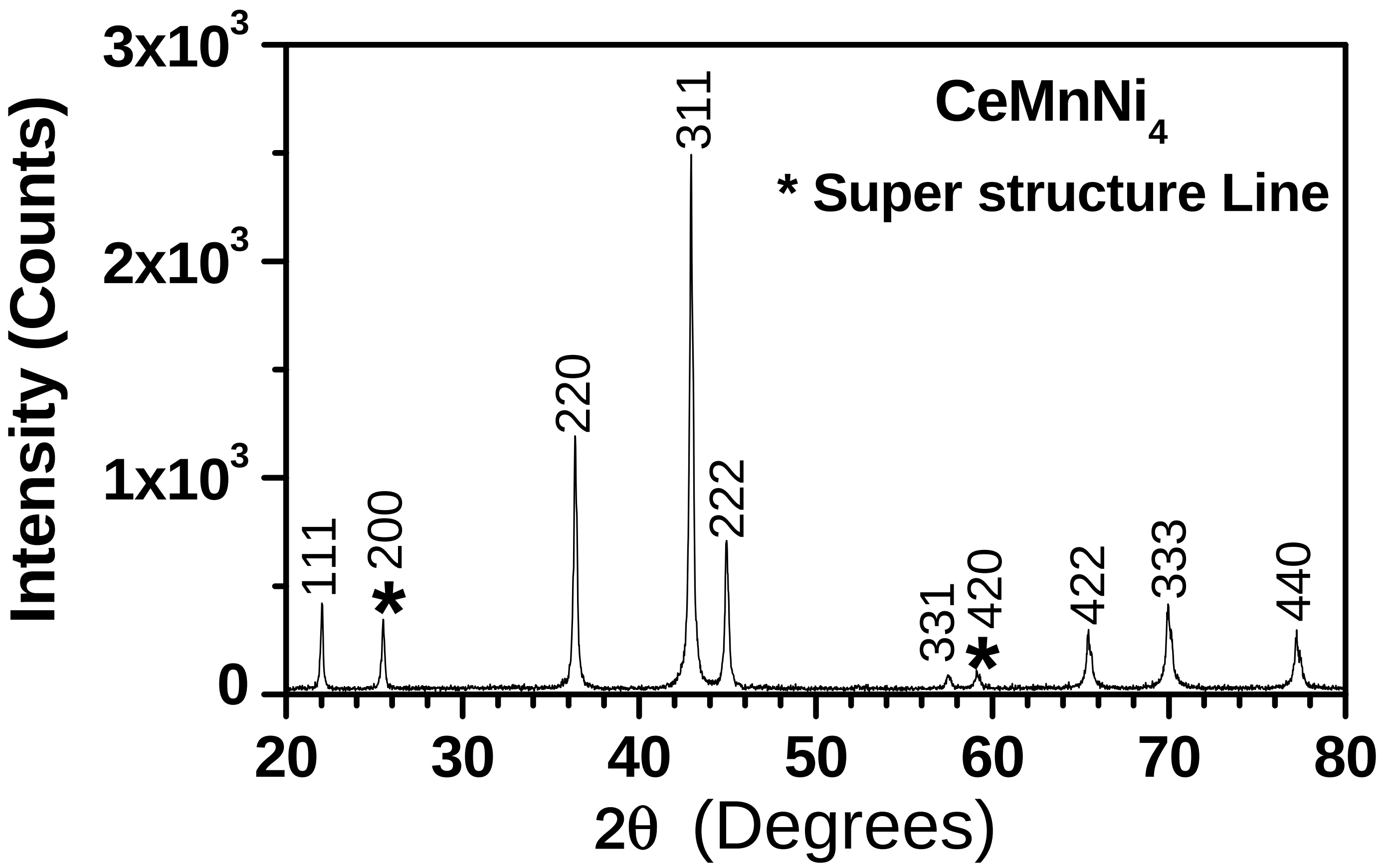

Figure 1 (S Singh et al.)

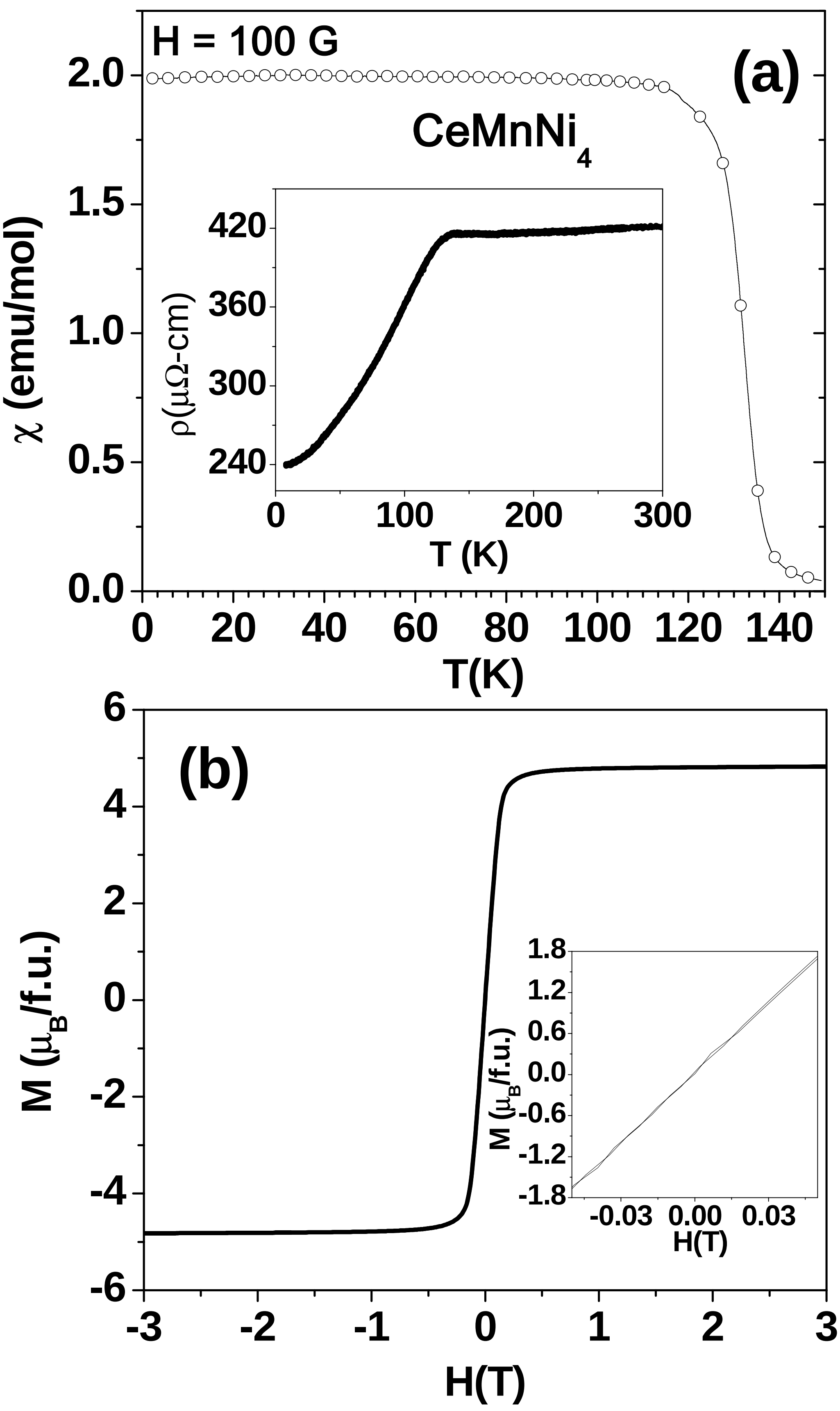

Figure 2 (Singh et al.)

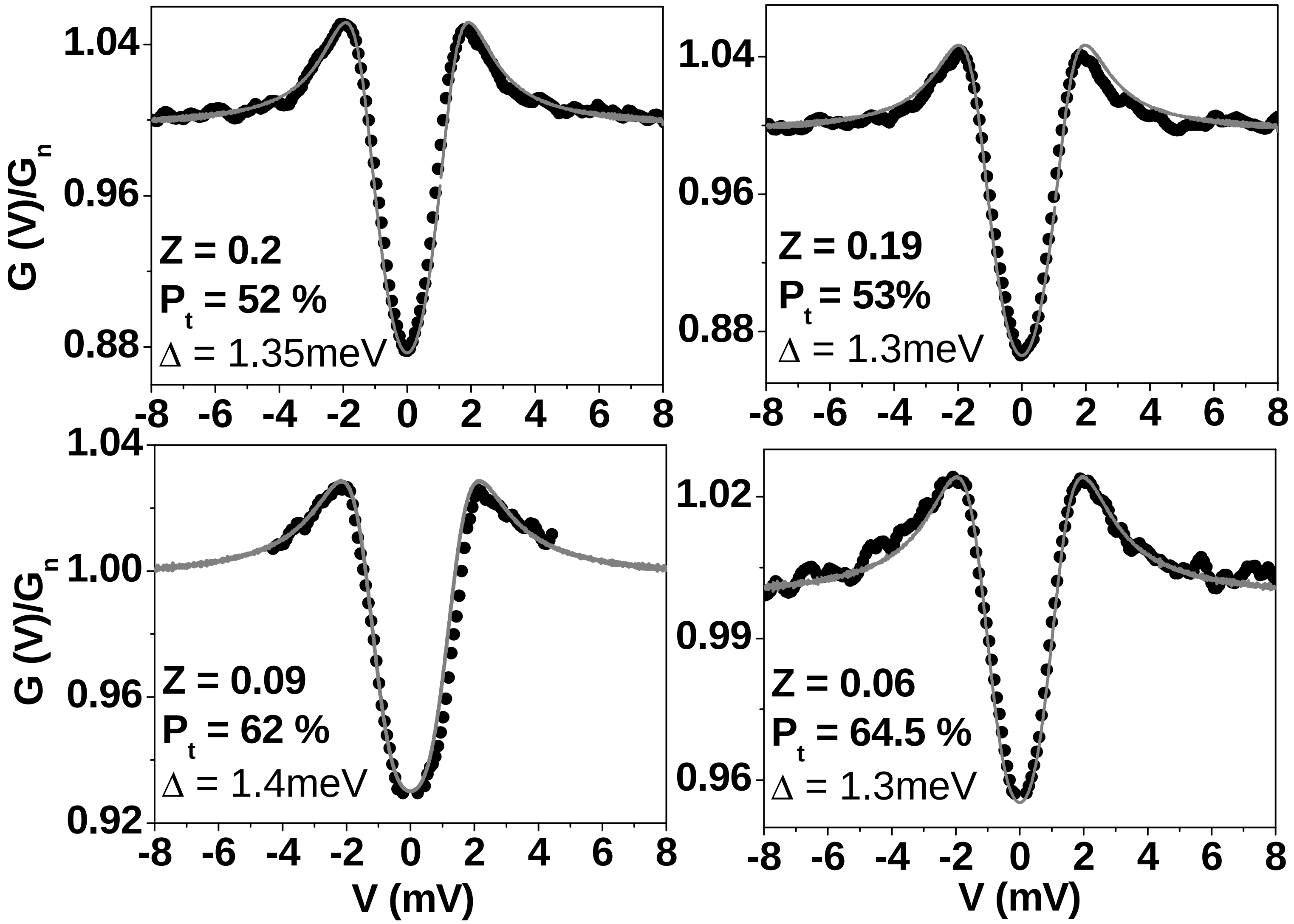

Figure 3 (Singh et al.)

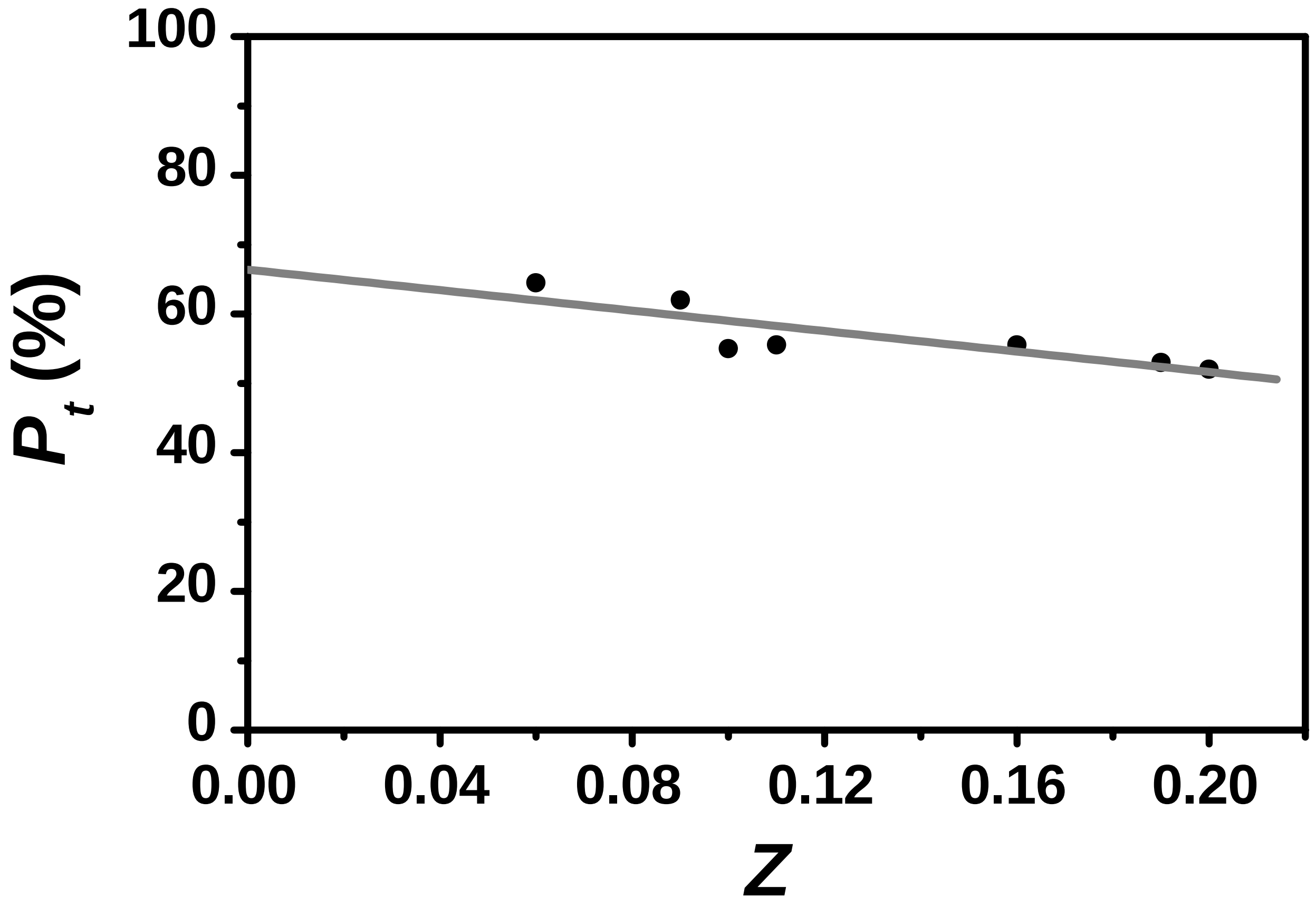

Figure 4 (S Singh et al)